\documentstyle[twoside, epsf]{article}

\input ibvs2.sty

\begin{document}
\IBVShead{5661}{25 November 2005}

\IBVStitletl{The rapid fading of V1647\,Orionis:}{the sudden end of a
FUor-type eruption?}

\IBVSauth{K\'osp\'al, \'Agnes$^1$; \'Abrah\'am, P\'eter$^{1}$,
  Acosta-Pulido, Jos\'e$^{2}$; Csizmadia, Szil\'ard$^1$;
  Eredics, M\'aria$^{1}$; Kun, M\'aria$^{1}$; R\'acz, Mikl\'os$^{1}$}

\IBVSinst{Konkoly Observatory, 1525 Budapest, P.O.Box 67, Hungary,
  e-mail: kospal@konkoly.hu}
\IBVSinst{Instituto de Astrof\'\i{}sica de Canarias, 38205 La Laguna,
  Tenerife, Canary Islands, Spain}

\SIMBADobj{V1647 Ori}
\IBVStyp{eruptive young stellar object}
\IBVSkey{photometry}
\IBVSabs{The young eruptive star V1647\,Ori started a rapid fading in}
\IBVSabs{October 2005 and now is only 1 mag above the pre-outburst level.}
\IBVSabs{Colour changes indicate that the fading is not due to increasing}
\IBVSabs{extinction.}

\begintext

V1647\,Orionis ($\alpha_{2000}$ = 05\hr 46\mm 13\fsec13,
$\delta_{2000}$ = $-$00\deg 06\arcm 04\farcs8) is a young eruptive
star, which went into outburst in November--December 2003. The star
brightened by about 4 mag in the I$_C$ band in 4 months, suggesting
that we witness either an FU\,Orionis-type (FUor) or an EX\,Lupi-type
(EXor) outburst. Since then, the object is gradually fading at both
optical (BVRI) and near-infrared (JHK) wavelengths with a rate which
is rather typical of FUors (see photometric measurements of Brice\~no
et al.~2004, Maheswar \& Bhatt 2004, Masi et al.~2004, McGehee et
al.~2004, Ojha et al.~2004, Semkov 2004, and Walter et al.~2004).

In this paper we present observations of V1647\,Ori using the 1m RCC
telescope of the Konkoly Observatory (Hungary) equipped with Cousins
V(RI)$_{C}$ filters and a Princeton VersArray:1300B CCD camera (image
scale: 0\farcs3, field of view: 6\farcm8 $\times$
6\farcm6). Integration time was selected so that the comparison stars
would not be saturated. This resulted in integration times between 180
and 600 s. With each filter, 3--10 frames were taken. All frames were
bias-subtracted and flat-fielded and were corrected for cosmic rays.
In the case of V and R$_C$ filters, the images were shifted and
co-added, and photometry was done in the single co-added image. In the
case of I$_C$ filter, photometry was done on each individual frame,
and the resulting magnitudes were averaged.

Photometry was performed using IRAF in the following way: on each
(co-added in V and R$_C$, individual in I$_C$) frame, 4 to 6 isolated,
non-saturated stars were selected to build the PSF. Then,
PSF-photometry was obtained for V1647\,Ori and for 4 comparison stars
(denoted as `A', `B', `C' and `G' by Semkov 2004). Instrumental
magnitude differences between V1647\,Ori and the comparison stars were
transformed to the standard Cousins-system, using the standard
magnitudes of comparison stars given by Semkov (2004), and Henden
(2004). The resulting magnitudes are presented in Tab.~1. In the case
of V and R$_C$ filters, the errors come from the uncertainties of the
standard transformation and from the formal errors of the photometry
given by IRAF, while in the case of I$_C$ filter, errors are
dominated by uncertainties of the standard transformation and the
scatter of the individual magnitudes.

\begin{table}
\centerline{Table 1.~Photometry of V1647\,Ori in October--November 2005.}
\begin{center}
\begin{tabular}{ccccc}
\hline
Date        & JD $-$ 2,453,000 & V                & R$_C$            & I$_C$            \\
\hline \hline
04 Oct 2005 & 648.54           & 20.45 $\pm$ 0.10 & 18.70 $\pm$ 0.10 & 16.31 $\pm$ 0.08 \\
05 Oct 2005 & 649.53           & 20.48 $\pm$ 0.10 & 18.55 $\pm$ 0.04 & 16.21 $\pm$ 0.07 \\
09 Oct 2005 & 653.64           & $-$              & $-$              & 16.42 $\pm$ 0.07 \\
10 Oct 2005 & 654.64           & $-$              & $-$              & 16.20 $\pm$ 0.10 \\
15 Oct 2005 & 659.65           & $-$              & $-$              & 16.33 $\pm$ 0.02 \\
19 Oct 2005 & 663.59           & $-$              & $-$              & 16.36 $\pm$ 0.05 \\
28 Oct 2005 & 672.55           & 21.34 $\pm$ 0.20 & 19.51 $\pm$ 0.10 & 17.13 $\pm$ 0.10 \\
30 Oct 2005 & 674.58           & 21.55 $\pm$ 0.15 & 19.68 $\pm$ 0.05 & 17.25 $\pm$ 0.08 \\
31 Oct 2005 & 675.57           & 21.74 $\pm$ 0.10 & 19.83 $\pm$ 0.04 & 17.44 $\pm$ 0.07 \\
17 Nov 2005 & 692.49           & $-$              & $-$              & 17.67 $\pm$ 0.08 \\
19 Nov 2005 & 694.60           & $-$              & $-$              & 17.80 $\pm$ 0.10 \\
\hline
\end{tabular}
\end{center}
\end{table}

In Fig.~1 we plotted the I$_C$ light curve of V1647\,Ori in
October--November 2005 complemented with some of our measurements from
February--March 2004 (these data were taken with the same instrument
and reduced with the same method as in October--November 2005, and
belong to a more comprehensive study, Acosta-Pulido et al., in prep.)
We also plotted data points from Brice\~no et al.~(2004), who measured
the brightening of the star. The overlapping points (in February-March
2004) show that although magnitudes were calculated differently
(Brice\~no et al.~used aperture photometry, we used PSF-photometry),
the values agree well, thus the comparison of the two datasets is
justifiable.

V1647\,Ori reached its peak brightness in February 2004, and faded by
approximately 1.5 mag by October 2005. Then, between October and
November 2005 the star suddenly dropped by more than 1 mag. Due to
this sudden, rapid fading, V1647\,Ori now is only 1~mag above the
pre-outburst level. This means that the present fading rate is
1~mag/month, as opposed to 0.1 mag/month in 2004 (calculated from the
data of Semkov 2004 or Walter et al.~2004).

In order to check whether the fading is caused by increasing
extinction, we plotted our measurements on a colour-colour diagram
(Fig.~2). The standard reddening path (Cohen et al.~1981) is also
shown. For comparison, we also plotted data points from McGehee et
al.~(2004) who measured V1647\,Ori close to peak brightness in
V(RI)$_C$, and performed PSF-photometry similarly to us. From Fig.~2
one can conclude that
\begin{itemize}
\item no significant colour change can be seen
   during the rapid fading in October-November 2005 (filled dots),
\item there is a significant colour change between the new
   measurements (filled dots) and those close to peak brightness (open
   squares). This colour change cannot fully be explained by
   increasing extinction, since the colour variations do not follow
   the reddening path. Thus, the observed colour changes are at least
   partly intrinsic. 
\end{itemize}

Supposing that the fading rate remains unchanged, the star will return
to the pre-outburst state by mid-December 2005. If this prediction
holds true, then the total length of the outburst of V1647\,Ori is 2
years, which makes it a unique (somewhat intermediate) object among
FUors and EXors.

\IBVSfig{9cm}{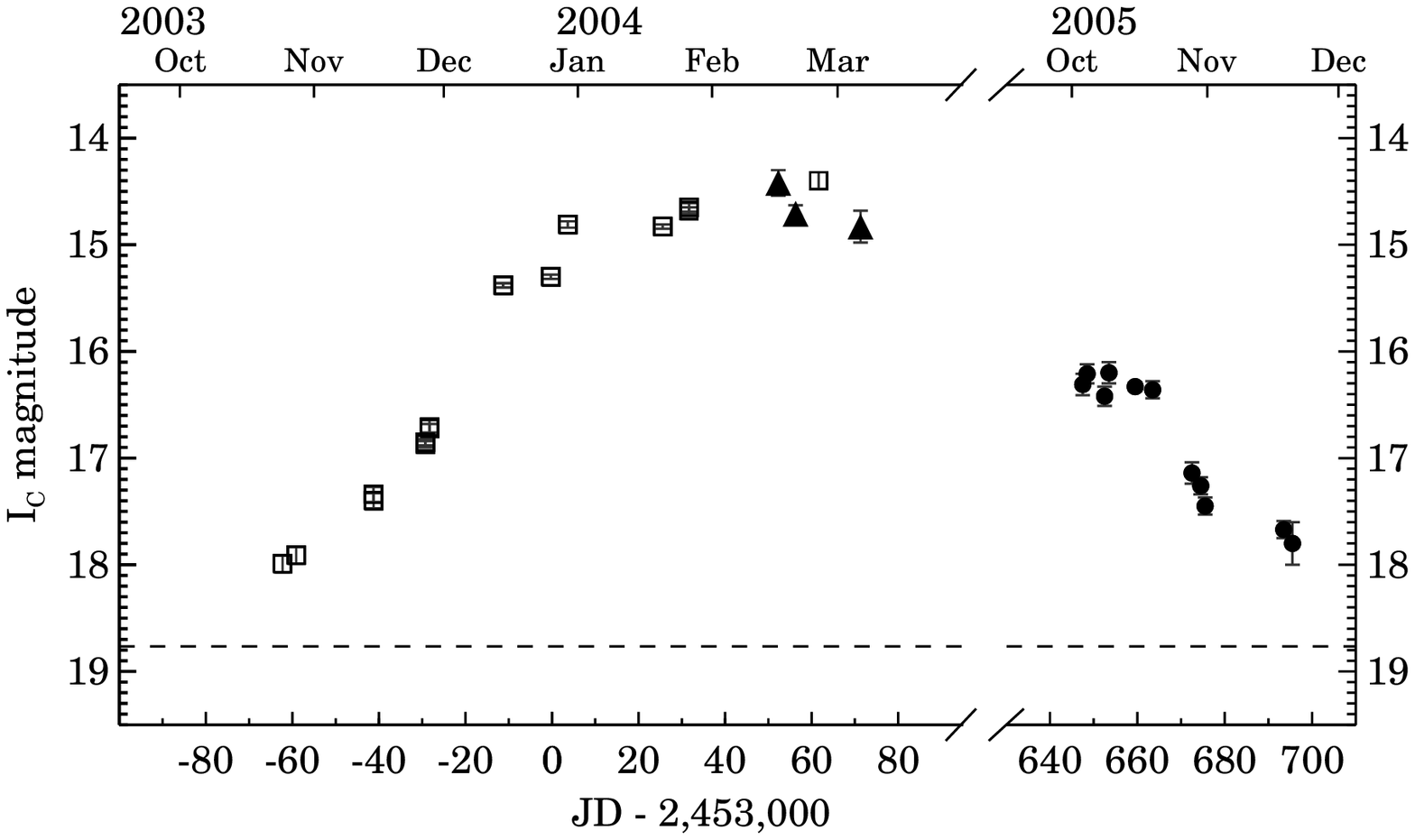}{I$_C$ light curve of V1647\,Ori. Filled
dots: photometry presented in this paper; filled triangles: photometry
taken with the same instrument and reduced with the same method as the
filled dots (Acosta-Pulido et al., in prep.); open squares: data from
Brice\~no et al.~2004. Dashed line indicates pre-outburst brightness
level.}

\IBVSfig{9cm}{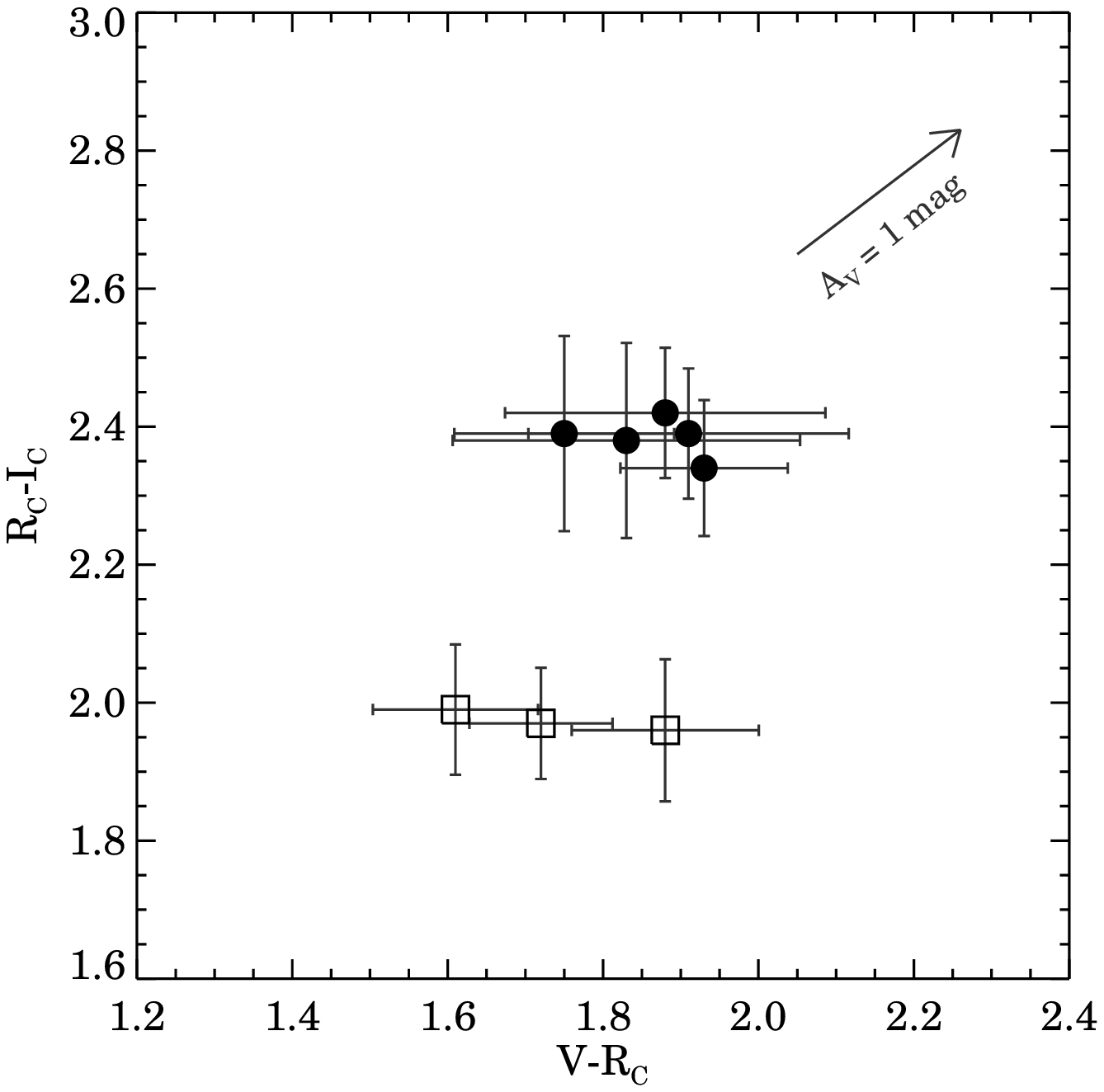}{Colour-colour diagram of
V1647\,Ori. Filled dots: our measurements obtained between 4~October
and 31 October 2005; open squares: data taken in February--April 2004
by McGehee et al.~2004. The line represents reddening path
corresponding to A$_V=1$.}

\textit{Acknowledgements}: We are grateful to J.~Jurcsik for her kind
help in the data analysis. The work was partly supported by the grants
OTKA\,T\,037508 and OTKA\,T\,049082 of the Hungarian Scientific
Research Fund.

\references 

Acosta-Pulido, J., Kun, M., \'Abrah\'am, P., K\'osp\'al, \'A.,
Csizmadia, Sz., Kiss, L.L., Mo\'or, A., Szabados, L., Benk\H{o}, J.,
Charcos-Llorens, M., Eredics, M., Kiss, Z.T., Manchado, A., R\'acz,
M., Ramos Almeida, C., Sz\'ekely, P., Vidal-N\'u\~nez, M.J., 2006,
{\it in prep.}

Brice\~no, C., Vivas, A.K., Hern\'andez, J., Calvet, N., Hartmann, L.,
Megeath, T., Berlind, P., Calkins, M., Hoyer, S, 2004, {\it ApJ}, {\bf
606}, L123

Cohen, J.G., Frogel, J.A., Persson, S.E., Elias, J.H., 1981, {\it
  ApJ}, {\bf 249}, 481

Henden, A., 2004, http://spiff.rit.edu/classes/phys440/lectures/new\_star/mcneil.dat

Maheswar, G., Bhatt, H. C., 2004, {\it IAU Circ.}, {\bf 8295}, 3

Masi, G., Mallia, F., Hornoch, K., Croman, R., Halderman, M., di
Cicco, D., Kreimer, E., 2004, {\it IAU Circ.}, {\bf 8290}, 2

McGehee, P.M., Smith, J.A., Henden, A.A., Richmond, M.W., Knapp, G.R.,
Finkbeiner, D.P., Ivezi\'c, \v{Z}., Brinkmann, J., 2004, {\it ApJ},
{\bf 616}, 1058

Ojha, D. K., Kusakabe, N., Tamura, M., 2004, {\it IAU Circ.}, {\bf
8306}, 2

Semkov, E. H., 2004, {\it IBVS}, {\bf 5578}

Walter, F.M., Stringfellow, G.S., Sherry, W.H., Field-Pollatou, A.,
  2004, {\it AJ}, {\bf 128}, 1872

\end{document}